\documentclass[11pt]{article}
\usepackage{geometry}                
\usepackage{graphicx}
\usepackage{amssymb}
\usepackage{epstopdf}
\usepackage{amsmath}
\title{Singularity dynamics: \\ Action and Reaction}
\author{M. Mazilu \\
School of Physics and Astronomy, \\
University of St. Andrews\\ 
St. Andrews, KY16 9SS, UK}%

\date{}                                           

\begin{document}
\maketitle

\abstract{The interaction between singular and regular fields is considered for Lorentz-invariant scalar and vector wave equations. The singular field is generated by a Dirac source term.  Its dynamics are deduced from the total field Lagrangian.  At non-relativistic speeds, the resulting equations of motion are those of a mass in a scalar potential. Using this method we deduce the relationship between source amplitude (scalar gravitational mass) and dynamic mass (inertial mass). Generalising this method implies Lorentz forces for charge singularities in the electromagnetic field and describes the dynamics and interaction of hypothetical magnetic monopoles. }

\section{Introduction}

The method presented builds on the action principle which offers a general tool to describe both particle and field dynamics. This principle represents one of the great unification in physics and applies equally well to Newtonian mechanics \cite{landau1}, general relativity and electromagnetic fields\cite{landau2}. Using this single principle it is possible to deduce the Euler-Lagrange equations describing the evolution of one system or of many interacting systems. However describing these interactions requires the assumption of an additional action term. In this paper, we show that it is possible to eliminate this additional assumption and deduce the reaction dynamics of the singularity. In the case of scalar field singularities we derive Newton's equation of motion together with the equivalence between Newtonian gravitational mass and inertial mass. For charges and magnetic monopoles we deduce their associated Lorentz force. 

Singularities, vortices and topological defects have received considerable attention over the years; they determine the properties of many interesting materials such as vortices in superconductors \cite{jensen1996}, superfluids \cite{duan1993} and two-dimensional condensate \cite{habib2000}, dislocations and defects in solid \cite{mermin1979} and liquid crystals \cite{straley1974}, optical phase singularities and optical vortices \cite{dennis2004} and even topological defect in cosmology \cite{srivastava1999}. In this paper, we use singularities together with the field action principle to deduce their dynamical behaviour. The method presented here is general and can be applied to all of the fields mentioned above. 

For clarity reasons we treat first the simplest case possible that being a Dirac source singularity in the Lorentz-invariant scalar wave equation. This leads to equations of motion similar to Newtonian mechanics (section 2). The method is generalised for other singularities and other fields such as a charges (section 3) where the electromagnetic field implies the Lorentz force. In the last section, we apply this principle to deduce the equations of motion of hypothetical magnetic monopoles where the dynamics concurs with the one obtained through charge-monopole symmetry considerations \cite{comay2004}. 

\section{Scalar field singularity: Inertia}

Before considering the anatomy of the field singularity, let us start by deducing the Euler-Lagrange equation for a scalar mass-less field $ U$ described by the following action
\begin{equation}\label{action}
\mathcal{S}_U= \int   {\cal L}(U) \; d^3 x\; dt
=  \frac{1}{2}  \int  \frac{1}{c^2} (\partial_t U)^2-( \mbox{\boldmath $\nabla$} U )^2   \; d^3 x \;dt
 \end{equation}
where ${\cal L}(U) $ corresponds to the Lagrangian density, $ \mbox{\boldmath $\nabla$}=(\partial_x,\partial_y,\partial_z)$ to the nabla operator and $c$ to the speed of light. The action principle implies that   the scalar field $ U$ is solution of the standard homogenous wave equation 
\begin{equation}\label{waveeq}
\mbox{\boldmath $\nabla$} ^2 U   -\frac{1}{c^2} \partial_t^2   U =0
\end{equation}
where $\mbox{\boldmath $\nabla$} ^2$ stands for the Laplacian operator. The next step in the singularity dynamics theory is to define a singular source term that maintains the Lorentz-invariance of the supporting field equation. Starting from the wave equation (\ref{waveeq}), we can define the stationary singular field $U_s$ as the time independent solution in the presence of a Dirac source at the origin of the coordinate system
\begin{equation}\label{annexe1}
\mbox{\boldmath $\nabla$} ^2  U_s   -\frac{1}{c^2} \partial_t^2  U_s  = m_s  \delta({\bf r})
 \end{equation}
where $m_s$ corresponds for the source amplitude. To define the Lorentz-invariant singularity, we consider equation (\ref{annexe1}) in a reference frame having the constant velocity $-{\bf v}$ with respect to the stationary singularity. In this reference frame the singularity has the velocity $\bf v$. The transformation between the two reference frames is \cite{lorrain}:
\label{transL}
 \begin{eqnarray}
{\bf r'}_\parallel &=&  \gamma ({\bf r}_\parallel+{\bf v} t)\nonumber \\
{\bf r'}_\perp &=& {\bf r}_\perp \\
t'&=& \gamma\left(t +\frac{{\bf v}\cdot{\bf r}}{c^2}\right). \nonumber
 \end{eqnarray}
where $\gamma=(1-{\bf v}^2/c^2)^{-1/2}$ and the prime stands for the coordinates in the moving reference frame. The subscripts $\parallel$ and $\perp$ represent the vectorial components parallel and perpendicular to the velocity $\bf v$. Applying these transformations to equation (\ref{annexe1}) leads us to a constant velocity moving singularity. Beforehand, we remark that the singularity can be decomposed into two components, $\delta({\bf r})=\delta({\bf r}_\parallel) \delta({\bf r}_\perp)$. Further, the wave equation and the scalar field $U_s$ is invariant with respect to the transformation that is it remains unchanged in the transition from one reference frame to another. It is only the singularity which changes in the moving reference system
\begin{eqnarray*}
\mbox{\boldmath $\nabla'$} ^2  U_s   -\frac{1}{c^2} \partial_{t'}^2  U_s
&=& m_s  \delta(\gamma ({\bf r'}_\parallel-{\bf v} t')) \delta({\bf r'}_\perp) \\
&=& m_s  \sqrt{1-\frac{\bf v^2}{c^2}}\delta({\bf r'}_\parallel-{\bf v} t')) \delta({\bf r'} _\perp)\\
&=& m_s  \sqrt{1-\frac{\bf v^2}{c^2}}\delta({\bf r'}-{\bf r'}_s)  
\end{eqnarray*}
where ${\bf r'}_s={\bf v} t'$ is the position of the singularity in the moving reference system. We have used the scaling property of the single variable Dirac distribution $|a|\delta(a x)=\delta(x)$.

We can now write the equation of a moving singularity on an arbitrary path $ {\bf r}_s(t)$ with the velocity ${\bf v}=\dot  {\bf r}_s$. The singular field $U_s$ is solution of 
\begin{equation}\label{mov}
\mbox{\boldmath $\nabla$} ^2  U_s   -\frac{1}{c^2} \partial_t^2  U_s  = m_s \sqrt{1-{\bf v}^2/c^2} \delta({\bf r}-{\bf r}_s).
\end{equation}
and  corresponds to a retarded potential. 

In the following, we are considering the action principle in the case of the linear superposition of the   singular field as defined in equation (\ref{mov}) and a given external field $U_0$ solution of (\ref{waveeq}). We remark here that the only free parameter of the system is the position of the Dirac singularity and its velocity. The Euler-Lagrange equation of motion is deduced when the action is stationary with respect path of the singularity.  The total action integral may be expressed as
\begin{eqnarray*}\label{action2}
\mathcal{S}_{U_0+U_s}({\bf r}_s(t))&= &\int  d^3 x\;dt\; {\cal L}(U_0+U_s)\nonumber \\
& =&   \frac{1}{2}  \int  d^3 x\;dt\;  \frac{1}{c^2} (\partial_t U_0)^2-( \mbox{\boldmath $\nabla$} U_0 )^2 +  \frac{1}{c^2} (\partial_t U_s)^2-( \mbox{\boldmath $\nabla$} U_s )^2 \nonumber \\
&& +\int  d^3 x\;dt\;   \frac{1}{c^2} (\partial_t U_0) (\partial_t U_s)-( \mbox{\boldmath $\nabla$} U_0 )( \mbox{\boldmath $\nabla$} U_s ) .
\end{eqnarray*}
This action integral can be decomposed into two parts. The first integral has no cross-terms between the external field and the singular field. This integral is independent with respect to the position of the singularity and its variation is zero.  This leaves only the second integral $\Delta\mathcal{S}_s$ for the action principle 
 \begin{eqnarray*}
\Delta\mathcal{S}_s&= & \int  d^3 x\;dt\; U_0\mbox{\boldmath $\nabla$}^2 U_s   - \frac{1}{c^2}  U_0\partial^2_t U_s  
\end{eqnarray*}
where we have integrated the action by parts and assumed the singular field to be zero on the boundary of the integration region. Next, we use equation (\ref{mov}) to replace the singular field by the Dirac singularity and finally we integrate in Lagrangian density in the spacial domain.
 \begin{eqnarray*}
\Delta\mathcal{S}_s
 &=& \int  d^3 x\;dt\;  U_0  m_s \sqrt{1-{\bf v}^2/^2} \delta({\bf r}-{\bf r}_s) \nonumber\\
&=& \int dt \; U_0  m_s \sqrt{1-{\bf v}^2/c^2} 
 \end{eqnarray*}
The equations of movement can be obtained by using the conventional particle Euler-Lagrange differential equation where the position and velocity correspond to those of the singularity. Indeed we have
\begin{eqnarray}
\frac{d}{dt} m_i {\bf v} &=& - m_g   \mbox{\boldmath $\nabla$}  U_0  \label{motion1}  
\end{eqnarray}
where we have defined an effective inertial $m_i$ and gravitational $m_g$ mass
\begin{equation}\label{mimg}\begin{split}
m_i&= \frac{U_0 m_s }{c^2\sqrt{1-{\bf v}^2/c^2}}\\
m_g&=m_s \sqrt{1-{\bf v}^2/c^2}.
 \end{split}\end{equation}
 
 One can also define a differential relationship between the two masses by taking the scalar product between the velocity and equation (\ref{motion1}). This leads to a energy balance equation that generalises the famous equations $E=mc^2$ to the dynamic domain.
\begin{eqnarray}
\frac{d}{dt}  m_i c^2 &=& m_g   \frac{\partial}{\partial t}  U_0 \label{motion2} 
\end{eqnarray}
Indeed, the energy balance equation shows that the variation of the inertial mass is proportional to its potential energy variation. 

Finally, we can deduce Newtons equation of motion for non-relativistic velocities and far from any other singularities
\begin{eqnarray}
\frac{d}{dt} m_s {\bf v} &=& 0 
\end{eqnarray}
where the scalar field gauge is chosen to be $U_0(\infty)=c^2$. 

\section{Charge singularity: Lorentz force}

In order to deduce the laws of motion of a charge we have to consider a singularity of the electromagnetic field described by Maxwell's equations. We proceed in a similar way to the scalar singularity case. In a first instance, we deduce the homogenous Maxwell's equations using the action principle. In a second step, we introduce a singular source to these equations making sure that their symmetries are maintained (i.e. Lorentz-invariance). Then, we write down the Lagrangian integral of the total field composed of an external field and the singular field. Finally, the action principle is applied to the total Lagrangian integral and the dynamics of the singularity is deduced. 

The electromagnetic field action is expressed as: 
\begin{equation}\label{action2em}
\mathcal{S}_{em}= \int  \left( \frac{\epsilon_0}{2}{\bf E}\cdot  {\bf E} - \frac{1}{2\mu_0} {\bf B}\cdot {\bf B} \right)\; d^3 x \; dt 
 \end{equation}
 where the electric and magnetic fields are defined by
 \begin{equation} \label{pot0}\begin{split}
 {\bf E}&=-\mbox{\boldmath $\nabla$}  V -\partial_t {\bf A} \\ 
 {\bf B}&=\mbox{\boldmath $\nabla$} \times {\bf A}\\
 0&= \mbox{\boldmath $\nabla$} \cdot {\bf A}+\epsilon_0\mu_0 \partial_t V 
  \end{split}\end{equation}
 and where $V$ and $\bf A$ are the electric scalar and magnetic vector potential respectively. The vacuum permittivity and permeability  is given by $\epsilon_0$ and $\mu_0$. The third equation corresponds to Lorentz gauge. Applying the action principle with respect to the potentials lets us deduce Maxwell's equations in absence of charges:
\begin{subequations} \label{maxq0}\begin{eqnarray}
-\mbox{\boldmath $\nabla$} \times {\bf E}- \partial_t {\bf B}&=&0  \label{max1a}\\
\frac{1}{\mu_0} \mbox{\boldmath $\nabla$} \times {\bf B}-\epsilon_0\partial_t {\bf E}&=& 0 \label{max1b}\\
\mbox{\boldmath $\nabla$} \cdot {\epsilon_0 \bf E}&=&0  \label{max1c}\\
 \mbox{\boldmath $\nabla$} \cdot {\bf B}&=&0.\label{max1d}
 \end{eqnarray}\end{subequations}

Like in the scalar case, we consider a single stationary Dirac singularity $q \delta({\bf r})$ in equation of the divergence of the electrical field (\ref{max1c}). The effect on the electromagnetic field of this source is equivalent to a  stationary point charge $q$ at the origin. The singular charge fields ${\bf E_q}$ and ${\bf B_q}$ are solutions of the in-homogenous Maxwell's equations:
\begin{equation}\label{onesing}
 \begin{split}
-\mbox{\boldmath $\nabla$} \times {\bf E_q} -\partial_t {\bf B_q} &= 0 \\
\frac{1}{\mu_0} \mbox{\boldmath $\nabla$} \times {\bf B_q} -\epsilon_0\partial_t {\bf E_q}  &=0 \\
\mbox{\boldmath $\nabla$} \cdot {\epsilon_0 \bf E_q}&=q \delta({\bf r})  \\
 \mbox{\boldmath $\nabla$} \cdot {\bf B_q}&=0
 \end{split}\end{equation}

The main difference with the scalar case is in that the potential fields $V_q$ and ${\bf A_q}$ change in the transition from one reference frame to another. Indeed, using the moving frame defined by (\ref{transL}) we have for the potentials:
 \begin{equation}\label{transA}
 \begin{split}
{\bf A_q'}_\parallel &=  \gamma \left({\bf A_q}_\parallel+\frac{ {\bf v} V_q}{c^2}\right)\\
{\bf A_q'}_\perp &= {\bf A_q}_\perp \\
V_q'&= \gamma (V_q +{\bf v}\cdot{\bf A_q})
 \end{split}
\end{equation}
where the speed of light is given by $1=c^2 \epsilon_0  \mu_0$.
Combining the  transformations (\ref{transL}) and (\ref{transA}) together with the definition of the electric and magnetic fields (\ref{pot0}) we can deduce the transformation relationship between these fields
 \begin{equation}\label{transB}
 \begin{split}
{\bf E_q'}_\parallel &=  {\bf E_q}_\parallel\\
{\bf E_q'}_\perp&= \gamma ({\bf E_q}_\perp-{\bf v}\times {\bf B_q})\\
{\bf B_q'}_\parallel &= {\bf B_q}_\parallel\\
{\bf B_q'}_\perp&= \gamma \left({\bf B_q}_\perp+\frac{ {\bf v}\times {\bf E_q}}{c^2}\right).
 \end{split}
\end{equation}
This implies the following transformation for equations (\ref{onesing}):
 \begin{eqnarray*}
-\mbox{\boldmath $\nabla'$} \times {\bf E_q'} -\partial_{t'} {\bf B_q'} &=& 0\\
\frac{1}{\mu_0} \mbox{\boldmath $\nabla'$} \times {\bf B_q'} -\epsilon_0\partial_{t'} {\bf E_q'}  &=& q {\bf v} \gamma\delta(\gamma ({\bf r'}_\parallel-{\bf v} t')) \delta({\bf r'}_\perp) \\
\mbox{\boldmath $\nabla'$} \cdot {\epsilon_0 \bf E_q'}&=&q  \gamma\delta(\gamma ({\bf r'}_\parallel-{\bf v} t')) \delta({\bf r'}_\perp)  \\
 \mbox{\boldmath $\nabla'$} \cdot {\bf B_q'}&=&0
 \end{eqnarray*}
 
Applying the scaling property of the single variable Dirac distribution $|a|\delta(a x)=\delta(x)$ and introducing the definition  of the position of the singularity  ${\bf r'}_q={\bf v} t'$ in the moving reference system we get:
 \begin{eqnarray*}
\mbox{\boldmath $\nabla'$} \times {\bf E_q'} +\partial_{t'} {\bf B_q'} &=& 0\\
\frac{1}{\mu_0} \mbox{\boldmath $\nabla'$} \times {\bf B_q'} -\epsilon_0\partial_{t'} {\bf E_q'}  &=& q {\bf v} \delta({\bf r'}-{\bf r'}_q) \\
\mbox{\boldmath $\nabla'$} \cdot {\epsilon_0 \bf E_q'}&=&q \delta({\bf r'}-{\bf r'}_q)  \\
 \mbox{\boldmath $\nabla'$} \cdot {\bf B_q'}&=&0
 \end{eqnarray*}
where the relativistic coefficient $\gamma$ in the source terms cancelled out. The reason for this is the covariant behaviour of the electromagnetic potentials as opposed to the invariant scalar potential in equation (\ref{mov}). 

This procedure gives us the form of the singularity following an arbitrary path ${\bf r}_q(t)$ with the velocity ${\bf v}=\dot  {\bf r}_q$.  The resulting vector singularity is equivalent to a charge and its associated current. The singular charge fields ${\bf E}_q$ and ${\bf B}_q$ are solutions of the in-homogenous Maxwell's equations:
\begin{equation}\label{maxq}
 \begin{split}
-\mbox{\boldmath $\nabla$} \times {\bf E_q} -\partial_t {\bf B_q} &= 0 \\
\frac{1}{\mu_0} \mbox{\boldmath $\nabla$} \times {\bf B_q}-\epsilon_0\partial_t {\bf E _q} &=q  {\bf v} \delta({\bf r}-{\bf r}_q) \\
\mbox{\boldmath $\nabla$} \cdot {\epsilon_0 \bf E_q}&=q \delta({\bf r}-{\bf r}_q)  \\
 \mbox{\boldmath $\nabla$} \cdot {\bf B_q}&=0.
 \end{split}\end{equation}
 
To deduce the equations of motion of this moving charge singularity we define the total field as the linear superposition of the singular fields ${\bf E_q}$ and ${\bf B_q}$ and the external fields ${\bf E_0}$ and ${\bf B_0}$:
 \begin{eqnarray*}
{\bf E}&=&  {\bf E_q} + {\bf E_0}\\
{\bf B}&=&  {\bf B_q} + {\bf B_0}
 \end{eqnarray*}
 where the external fields are decomposed onto the external scalar and vector potentials
\begin{eqnarray*}
 {\bf E_0}&=&-\mbox{\boldmath $\nabla$}  V_0 -\partial_t {\bf A_0} \\ 
 {\bf B_0}&=&\mbox{\boldmath $\nabla$} \times {\bf A_0}\\
 0&=& \mbox{\boldmath $\nabla$} \cdot {\bf A_0}+\epsilon_0\mu_0 \partial_t V_0.
  \end{eqnarray*}

Introducing the total fields ${\bf E}$ and ${\bf B}$  into the electromagnetic Lagrangian (\ref{action2em}) infers the following singularity  action:
\begin{equation}
\Delta\mathcal{S}_{sq}= \int  \left( \epsilon_0{\bf E_0}\cdot  {\bf E_q} - \frac{1}{\mu_0} {\bf B_0}\cdot {\bf B_q} \right)\; d^3 x \; dt 
 \end{equation}
where we have kept only the terms that vary with respect to the trajectory of the singularity. Indeed, the terms in ${\bf E_i} \cdot  {\bf E_i}$ and ${\bf B_i} \cdot  {\bf B_i}$, where $(i=q,0)$, are not dependent on the position of the charge singularity. Rewriting the external field with the help of the scalar $V_0$ and vector potential ${\bf A_0}$ allows us to integrate by parts:
\begin{eqnarray*}
\Delta\mathcal{S}_{sq}&=& \int  \left( \epsilon_0\left(-\mbox{\boldmath $\nabla$}  V_0 -\partial_t {\bf A_0} \right)\cdot  {\bf E_q} - \frac{1}{\mu_0} \left(\mbox{\boldmath $\nabla$} \times {\bf A_0}\right)\cdot {\bf B_q} \right)\; d^3 x \; dt \\
&=& \int  \left( \epsilon_0\left(V_0\mbox{\boldmath $\nabla$} \cdot {\bf E_q}  +{\bf A_0}\cdot\partial_t {\bf E_q}  \right) - \frac{1}{\mu_0} \left(\mbox{\boldmath $\nabla$} \times {\bf B_q}\right)\cdot {\bf A_0} \right)\; d^3 x \; dt
 \end{eqnarray*}
where we have assumed the fields ${\bf E_q}$ and ${\bf B_q}$ to be zero on the boundary of the integration domain. Using equations (\ref{maxq}) we can further simplify the action of the singularity
\begin{eqnarray*}
\Delta\mathcal{S}_{sq}
&=&  \int  \left(q V_0  \delta({\bf r}-{\bf r}_q)  - q {\bf A_0} \cdot {\bf v} \delta({\bf r}-{\bf r}_q) \right)\; d^3 x \; dt\\
&=&  \int  \left(q V_0 - q {\bf A_0} \cdot {\bf v} \right)\; dt
 \end{eqnarray*}
which leads directly to Lorentz forces for a charge in the external field defined by ${\bf E_0}$ and ${\bf B_0}$. To convince ourselves of this we can rigidly link a scalar mass singularity introduced in previous section to a charge singularity. The total action of this double singularity is:
 \begin{eqnarray}
\Delta\mathcal{S}_s+\Delta\mathcal{S}_{sq}
&=&  \int  \left(U_0 m_s \sqrt{1-\frac{{\bf v}^2}{c^2}}+q V_0 -q {\bf A_0} \cdot {\bf v} \right)\; dt
 \end{eqnarray}
which lead to the following equation of movement
\begin{eqnarray}
\frac{d}{dt} m_i {\bf v} &=& - m_g   \mbox{\boldmath $\nabla$}  U_0 + q {\bf E_0} +q {\bf v}\times {\bf B_0}   
\end{eqnarray}
where  $m_i$ and $m_g$ are the defined by (\ref{mimg}). The energy balance equation is 
\begin{eqnarray}
\frac{d}{dt}  m_i c^2 &=& m_g   \frac{\partial}{\partial t}  U_0+q {\bf v}\cdot{\bf E_0}.
\end{eqnarray}
 
 \section{Magnetic Monopole singularity} 
 
  Magnetic monopoles are hypothetical particles that behave in a similarly to charged particles with magnetic and electric fields inverted (\cite{comay2004}). The monopoles correspond to sources of the magnetic field in equation (\ref{max1d}). In the following we are deducing the equations of motion of magnetic source singularities by considering a Dirac source $q_m \delta({\bf r})$ in equations (\ref{max1d}) where $q_m$ corresponds to the monopole strength. 

Before proceeding further, we have to generalise the potential decomposition of the electromagnetic field. Indeed, the potentials that we used for the charged singularities does not allow the presence of sources of the magnetic field. Conventionally the magnetic field is by definition a rotational field. Therefore, we define a symmetric electromagnetic potential decomposition that includes an additional magnetic scalar and  electric vector potential. 
\begin{equation}\label{potqm}
 \begin{split}
{\bf E}&= -\mbox{\boldmath $\nabla$}  V_q -\partial_t {\bf A _q}+\frac{1}{\epsilon_0} \mbox{\boldmath $\nabla$} \times{\bf A_m}  \\
{\bf B}&=\mbox{\boldmath $\nabla$}\times {\bf A_q} +\mu_0\mbox{\boldmath $\nabla$}  V_m +\mu_0\partial_t {\bf A_m} \\
 0&= \mbox{\boldmath $\nabla$} \cdot {\bf A_q}+\epsilon_0\mu_0 \partial_t V_q \\
 0&= \mbox{\boldmath $\nabla$} \cdot {\bf A_m}+\epsilon_0\mu_0 \partial_t V_m 
\end{split}\end{equation}
where the subscripts $(q,m)$ stand respectively for the charge and monopole potentials or fields. Using all the newly defined potential definitions we can apply again the action principle the electromagnetic action defined by equation (\ref{action2em}). The resulting evolution equations are identical to equations (\ref{maxq0}). The only difference is the possibility to define a monopole source density in equation (\ref{max1d:4}). 
\begin{subequations} \label{maxm0}\begin{eqnarray}
- \mbox{\boldmath $\nabla$} \times {\bf E}- \partial_t {\bf B}&=&0  \\
\frac{1}{\mu_0} \mbox{\boldmath $\nabla$} \times {\bf B}-\epsilon_0\partial_t {\bf E}&=& 0 \\
\mbox{\boldmath $\nabla$} \cdot {\epsilon_0 \bf E}&=&0  \\
 \mbox{\boldmath $\nabla$} \cdot  {\bf B}&=&0. \label{max1d:4}
 \end{eqnarray}\end{subequations}
 
For the monopole singularity, we consider a single stationary Dirac singularity at the origin. The singularity constitutes the free term of equation (\ref{max1d:4}) and acts as a source of magnetic field. The singular fields $\bf E_{m}$ and $\bf B_{m}$ are solution of:
\begin{equation}\label{onesingm}
 \begin{split}
-\mbox{\boldmath $\nabla$} \times {\bf E_{m}} -\partial_t {\bf B_{m}} &= 0\\
\frac{1}{\mu_0} \mbox{\boldmath $\nabla$} \times {\bf B_{m}} -\epsilon_0\partial_t {\bf E_{m}}  &=0\\
\mbox{\boldmath $\nabla$} \cdot {\epsilon_0 \bf E_{m}}&=0 \\
 \mbox{\boldmath $\nabla$} \cdot {\bf B_{m}}&=q_m \delta({\bf r})
 \end{split}\end{equation}

In this case there are four potential fields $V_{sq}$, $V_{sm}$,  ${\bf A_{sq}}$ and ${\bf A_{sm}}$ and the electric and magnetic field is defined by equations (\ref{potqm}). Taking the same frame transformation  defined by (\ref{transL}) we have for the potentials:
 \begin{equation}\label{transA2}
 \begin{split}
{\bf A'_{sq}}_\parallel &=  \gamma ({\bf A_{sq}}_\parallel+\frac{{\bf v} V_{sq}}{c^2})\\
{\bf A'_{sq}}_\perp &= {\bf A_{sq}}_\perp \\
V'_{sq}&= \gamma (V_{sq} +{\bf v}\cdot{\bf A_{sq}})\\
{\bf A'_{sm}}_\parallel &=  \gamma ({\bf A_{sm}}_\parallel+\frac{{\bf v} V_{sm}}{c^2})\\
{\bf A'_{sm}}_\perp &= {\bf A_{sm}}_\perp \\
V'_{sm}&= \gamma (V_{sm} +{\bf v}\cdot{\bf A_{sm}}).
 \end{split}
\end{equation}
The implied relationship between the electromagnetic fields in the two frames is identical to the transformation for the charges only field.
 \begin{equation}\label{transB2}
 \begin{split}
{\bf E'_{m}}_\parallel &=  {\bf E_m}_\parallel\\
{\bf E'_m}_\perp&= \gamma ({\bf E_m}_\perp-{\bf v}\times {\bf B_m})\\
{\bf B'_m}_\parallel &= {\bf B_m}_\parallel\\
{\bf B'_m}_\perp&= \gamma ({\bf B_m}_\perp+\frac{{\bf v}\times {\bf E_m}}{c^2})
 \end{split}
\end{equation}

Combining all the above transformations implies the singularity in equations (\ref{onesingm}) to transform as:
 \begin{eqnarray*}
-\mbox{\boldmath $\nabla'$} \times {\bf E'_m} -\partial_{t'} {\bf B'_m} &=& q_m {\bf v} \gamma\delta(\gamma ({\bf r'}_\parallel-{\bf v} t')) \delta({\bf r'}_\perp) \\
\frac{1}{\mu_0} \mbox{\boldmath $\nabla'$} \times {\bf B'_m} -\epsilon_0\partial_{t'} {\bf E'_m}  &=& 0  \\
\mbox{\boldmath $\nabla'$} \cdot {\epsilon_0 \bf E'_m}&=&0 \\
 \mbox{\boldmath $\nabla'$} \cdot {\bf B'_m}&=&q_m  \gamma\delta(\gamma ({\bf r'}_\parallel-{\bf v} t')) \delta({\bf r'}_\perp) 
  \end{eqnarray*}
 
Applying the scaling property of the single variable Dirac distribution $|a|\delta(a x)=\delta(x)$ and introducing the definition  of the position of the singularity  ${\bf r'}_m={\bf v} t'$ in the moving reference system we get:
 \begin{eqnarray*}
-\mbox{\boldmath $\nabla'$} \times {\bf E'_m}-\partial_{t'} {\bf B'_m} &=& q_m {\bf v} \delta({\bf r'}-{\bf r'}_m)  \label{amax4a}\\
\frac{1}{\mu_0} \mbox{\boldmath $\nabla'$} \times {\bf B'_m} -\epsilon_0\partial_{t'} {\bf E'_m}  &=& 0 \label{amax4b}\\
\mbox{\boldmath $\nabla'$} \cdot {\epsilon_0 \bf E'_m}&=&0\label{amax4c}\\
 \mbox{\boldmath $\nabla'$} \cdot {\bf B'_m}&=&q_m \delta({\bf r'}-{\bf r'}_m)  \label{amax4d}
 \end{eqnarray*}
where the relativistic coefficient $\gamma$ in the source terms cancelled out just like in the charge singularity case. 

Consequently, the electromagnetic singular fields $\bf E_{m}$ and $\bf B_{m}$ are solution of 
\begin{equation}\label{maxmag}
 \begin{split}
-\epsilon_0 \mbox{\boldmath $\nabla$} \times {\bf E_m} -\frac{1}{\mu_0}\partial_t {\bf B_m} &= q_m  {\bf v} \delta({\bf r}-{\bf r}_m)\\
\frac{1}{\mu_0} \mbox{\boldmath $\nabla$} \times {\bf B _m}-\epsilon_0\partial_t {\bf E _m} &=0\\
\mbox{\boldmath $\nabla$} \cdot {\epsilon_0 \bf E_m}&= 0\\
 \mbox{\boldmath $\nabla$} \cdot {\frac{1}{\mu_0} \bf B_m}&=q_m \delta({\bf r}-{\bf r}_m).
 \end{split}\end{equation}
 for a magnetic monopole singularity on an arbitrary path ${\bf r}_m(t)$ with a  velocity $\bf v=\dot r_m$.  

The equation of motion of the monopole singularity is deduced by defining the  total field as the linear superposition of the singular fields ${\bf E}_m$ and ${\bf B}_m$ and the external fields ${\bf E_0}$ and ${\bf B_0}$:
 \begin{eqnarray*}
{\bf E}&=&  {\bf E_m} + {\bf E_0}\\
{\bf B}&=&  {\bf B_m} + {\bf B_0}
 \end{eqnarray*}
  where the external fields can be decomposed onto four potential fields $V_{0q}$, $V_{0m}$,  ${\bf A_{0q}}$ and ${\bf A_{0m}}$
  \begin{subequations} \label{pot0c}\begin{eqnarray}
    {\bf E_0}&=& -\mbox{\boldmath $\nabla$}  V_{0q} -\partial_t {\bf A _{0q}}+\frac{1}{\epsilon_0} \mbox{\boldmath $\nabla$} \times{\bf A_{0m}} \label{pot1} \\
{\bf B_0}&=&\mbox{\boldmath $\nabla$}\times {\bf A_{0q}} +\mu_0\mbox{\boldmath $\nabla$}  V_{0m} +\mu_0\partial_t {\bf A_{0m}} \label{pot2} \\
 0&=& \mbox{\boldmath $\nabla$} \cdot {\bf A_{0q}}+\epsilon_0\mu_0 \partial_t V_{0q} \\
 0&=& \mbox{\boldmath $\nabla$} \cdot {\bf A_{0m}}+\epsilon_0\mu_0 \partial_t V_{0m}.
    \end{eqnarray}\end{subequations}

Introducing the total field in the electromagnetic Lagrangian (eq. \ref{action2em}) infers the following singularity  action:
\begin{equation}
\Delta \mathcal{S}_{sm}= \int  \left( \epsilon_0{\bf E_0}\cdot  {\bf E_m} - \frac{1}{\mu_0} {\bf B_0}\cdot {\bf B_m} \right)\; d^3 x \; dt 
 \end{equation}
where we have kept only the terms that vary with respect to the trajectory of the singularity. Indeed, the terms in ${\bf E_i} \cdot  {\bf E_i}$ and ${\bf B_i} \cdot  {\bf B_i}$, where $(i=m,0)$, are not dependent on the position of the charge singularity. Rewriting the external field with the help of the potentials defined in equations (\ref{pot1}) and (\ref{pot2}) allows us to perform an integration by part
\begin{eqnarray*}
\Delta\mathcal{S}_{sm}&=& \int  \left( \epsilon_0\left(-\mbox{\boldmath $\nabla$}  V_{0q} -\partial_t {\bf A _{0q}}+\mbox{\boldmath $\nabla$} \times{\bf A_{0m}}  \right)\cdot  {\bf E_m} \right. \\ 
&&\left. - \frac{1}{\mu_0} \left(\mbox{\boldmath $\nabla$}\times {\bf A_{0q}} +\mbox{\boldmath $\nabla$}  V_{0m} +\partial_t {\bf A _{0m}}\right)\cdot {\bf B_m} \right)\; d^3 x \; dt \\
&=&\int  \left( \epsilon_0\left( V_{0q} \mbox{\boldmath $\nabla$} {\bf E_m} +{\bf A_{0q}}\partial_t {\bf E_m} +{\bf A_{0m}}\mbox{\boldmath $\nabla$} \times {\bf E_m}  \right) \right. \\ && \left.
- \frac{1}{\mu_0} \left( {\bf A_{0q}} \mbox{\boldmath $\nabla$}\times{\bf B_m} -V_{0m} \mbox{\boldmath $\nabla$}  {\bf B_m} -{\bf A _{0m}}\partial_t {\bf B_m} \right) \right)\; d^3 x \; dt
 \end{eqnarray*}
where we have assumed the fields ${\bf E_m}$ and ${\bf B_m}$ to be zero at infinity i.e. the boundary of integration. Using equations (\ref{maxmag}) we can further simplify the action of the singularity
\begin{eqnarray*}
\Delta\mathcal{S}_{sm}
&=&  \int  \left(q_mV_{0m}  \delta({\bf r}-{\bf r}_q)  - q_m {\bf A_{0m}} \cdot {\bf v} \delta({\bf r}-{\bf r}_q) \right)\; d^3 x \; dt\\
&=&  \int  \left(q_m V_{0m} - q_m {\bf A_{0m}} \cdot {\bf v} \right)\; dt
 \end{eqnarray*}
which leads directly to a Lorentz type forces. Like in the case of the charge singularity, we can rigidly link a scalar mass singularity introduced in previous section to a monopole singularity. The total action of the double singularity is:
 \begin{eqnarray}
\Delta\mathcal{S}_{s}+\Delta\mathcal{S}_{sm}
&=&  \int  \left(U_0 m_s \sqrt{1-\frac{{\bf v}^2}{c^2}}+q_m V_{0m} - q_m {\bf A_{0m}} \cdot {\bf v} \right)\; dt
 \end{eqnarray}
which lead to the following equation of movement
\begin{eqnarray}
\frac{d}{dt} m_i {\bf v} &=& - m_g   \mbox{\boldmath $\nabla$}  U_0 + q_m {\bf B_{0m}} +q_m {\bf v}\times {\bf E_{0m}}  
\end{eqnarray}
where  $m_i$ and $m_g$ are the defined by (\ref{mimg}) and 
  \begin{subequations} \label{potextm}
\begin{eqnarray*}
{\bf E_{0m}}&=& \frac{1}{\epsilon_0} \mbox{\boldmath $\nabla$} \times{\bf A_{0m}} \\
{\bf B_{0m}}&=&\mu_0\mbox{\boldmath $\nabla$}  V_{0m} +\mu_0\partial_t {\bf A_{0m}} 
\end{eqnarray*}\end{subequations}
correspond to the magnetic monopole part of the electromagnetic field. We remark here that the force acting on the monopole is only due to the monopole part of the electromagnetic field. Consequently, there is no direct interaction between charges and magnetic monopoles. 

The energy balance equation is 
\begin{eqnarray}
\frac{d}{dt}  m_i c^2 &=& m_g   \frac{\partial}{\partial t}  U_0  +q_m {\bf v}\cdot{\bf B_{0m}}.
\end{eqnarray}

\section{Conclusion}

The field generated by a singularity gives rise to the laws of motion of the singularity. In other words, the action of the singularity on the field implies its dynamic reaction to the field. We have shown this to be the case for neutral bodies, charges and hypothetical magnetic monopoles. Using our method we have deduced not only the dynamics of masses, but also the relationship between the "gravitational" source strength  and the associated inertial mass. The charge singularity was shown to be the origin of the Lorentz force. The magnetic monopoles have an equivalent Lorentz force. 

The conventional Lorentz force does not involve the magnetic monopole field.  Reciprocally, the monopole Lorentz force does not involve fields originating from charged particles. Consequently, charges and magnetic monopoles do not interact, making the discovery of monopoles using charges difficult. A major question to be addressed concerns the role played by electromagnetic radiation. Can it provide the link between charges and magnetic monopoles?

\end{document}